\documentclass[10pt]{revtex4}
\usepackage{amssymb}
\usepackage{latexsym}
\usepackage{epsfig}
\usepackage{amsmath}
\begin{document}                                          

\title{Interacting generalized ghost dark energy in a non-flat universe}

\author{Esmaeil Ebrahimi$^{1,3}$ \footnote{eebrahimi@uk.ac.ir}, Ahmad  Sheykhi$^{2,3}$ \footnote{
asheykhi@shirazu.ac.ir} and Hamzeh Alavirad
$^{4}$\footnote{hamzeh.alavirad@kit.edu}}
\address{$^1$ Department of Physics, Shahid Bahonar University, PO Box 76175, Kerman, Iran\\
          $^2$ Physics Department and Biruni Observatory, College of Sciences, Shiraz University, Shiraz 71454, Iran \\
          $^3$ Research Institute for Astronomy and Astrophysics of Maragha (RIAAM), P. O. Box 55134-441, Maragha,
          Iran\\
          $^4$ Institute for Theoretical Physics, Karlsruhe Institute of Technology (KIT), 76128 Karlsruhe, Germany}
\begin{abstract}
We investigate the generalized Quantum Chromodynamics (QCD) ghost
model of dark energy in the framework of Einstein gravity. First,
we study the non-interacting generalized ghost dark energy in a
flat Friedmann-Robertson-Walker (FRW) background. We obtain the
equation of state parameter, $w_D=p/\rho$, the deceleration
parameter, and the evolution equation of the generalized ghost
dark energy. We find that, in this case, $w_D$ cannot cross the
phantom line ($w_D>-1$) and eventually the universe approaches a
de-Sitter phase of expansion $(w_D\rightarrow-1)$. Then, we extend
the study to the interacting ghost dark energy in both a flat and
non-flat FRW universe. We find that the equation of state
parameter of the interacting generalized ghost dark energy can
cross the phantom line ($w_D<-1$) provided the parameters of the
model are chosen suitably. Finally, we constrain the  model
parameters by using the Markov Chain Monte Carlo (MCMC) method and
a combined
dataset of SNIa, CMB, BAO and X-ray gas mass fraction.
\end{abstract}

\maketitle

\section{Introduction}
The cosmological data from type Ia Supernova, Large Scale
Structure(LSS) and Cosmic Microwave Background (CMB) indicate that
our universe is currently accelerating \cite{Rie}. To explain such
an acceleration in the framework of standard cosmology, one is
required to introduce a new type of energy with a negative
pressure usually called ``dark energy" (DE) in the literature. A
great variety of DE scenarios have been proposed to explain the
acceleration of the universe's expansion. One can refer to
\cite{sami,li} for a review of DE models. On the other hand, many
people believe in a modification of gravity, seeking an
explanation for the late time acceleration. According to this idea
the acceleration will be a part of the universe's expansion and
does not need to invoke any kind of DE component. As examples of
this approach one can look at Refs.
\cite{dgp,deffayet,arkani1,arkani2,Nozari}. It is important to
note that the detection of gravitational waves should be the
ultimate test for general relativity or alternatively the
definitive endorsement for extended theories \cite{Cor}.

In most scenarios for DE, people usually need to consider a new
degree of freedom or a new parameter, in order to explain the
acceleration of the cosmic expansion (see e.g. \cite{sheykhi} and
references therein). However, it would be nice to resolve the DE
puzzle without presenting any new degree of freedom or any new
parameter in the theory. One of the successful and beautiful
theories of modern physics is QCD which describes the strong
interaction in nature. However, resolution of one of its
mysteries, the U(1) problem, has remained somewhat unsatisfying.
Veneziano ghost field explained the U(1) problem in QCD
\cite{witten}. Vacuum energy of the ghost field can be used to
explain the time-varying cosmological constant in a spacetime with
nontrivial topology, since the ghost field has no contribution to
the vacuum energy in the Minkowskian spacetime \cite{urban1}. The
energy density of the vacuum ghost field is proportional to
$\Lambda_{QCD}^3 H$, where $\Lambda_{QCD}$ is the QCD mass scale
and $H$ is the Hubble parameter \cite{ohta}. It is well-known that
the cosmological constant model of DE suffers the coincidence and
the fine tuning problems. However, with correct choice of
$\Lambda_{QCD}$, the ghost dark energy (GDE) model does not
encounter the fine tuning problem anymore \cite{urban1,ohta}.
Phenomenological implications of the GDE model were discussed in
\cite{caighost}. In \cite{shemov} GDE in a non-flat universe in
the presence of interaction between DE and dark matter was
explored. The instability of the GDE model against perturbations
was studied in \cite{ebrins}. It was argued that the perfect fluid
for GDE is classically unstable against perturbations. Other
features of the GDE model have been investigated in Refs.
\cite{shemoveb,shebagh,ebrbd,rozas,karami,khodam1,khodam2,Obs}.

In all the above references
(\cite{caighost,shemov,ebrins,shemoveb,shebagh,ebrbd,rozas,karami,khodam1,khodam2,Obs})
the GDE was assumed to have the energy density of the form
$\rho_D=\alpha H$, while, in general, the vacuum energy of the
Veneziano ghost field in QCD is of the form $H+O(H^2)$ \cite{Zhi}.
This indicates that in the previous works on the GDE model,  only
the leading term $H$ has been considered. Motivated by the
argument given in \cite{mag}, one may expect that the subleading
term $H^2$ in the GDE model might play a crucial role in the early
evolution of the universe, acting as the early DE.  It was shown
\cite{caighost2} that taking the second term into account can give
better agreement with observational data compared to the usual
GDE. Hereafter we call this model the generalized ghost dark
energy (GGDE) and our main task in this paper is to investigate
the main properties of this model. In this model the energy
density is written in the form $\rho_D= \alpha H+\beta H^2$, where
$\beta$ is a constant.

In addition to the DE component, there is also another unknown
component of energy in our universe called "dark matter" (DM).
Since the nature of these two dark components are still a mystery
and they seem to have different gravitational behaviour, people
usually consider them separately and take their evolution
independent of each other. However, there exist observational
evidence of signatures of interaction between the two dark
components \cite{interact1,oli}.

On the other hand, based on the cosmological principle the
universe has three distinct geometries, namely open, flat and
closed geometry corresponding to $k=-1,0,+1$, respectively. For a
long time it was a general belief that the universe has a flat
($k=0$) geometry, mainly based on the inflation theory
\cite{guth}. With the development of observational techniques
people found deviations from the flat geometry \cite{spe}. For
example, CMB experiments \cite{Sie}, supernova measurements
\cite{Caldwell}, and WMAP data \cite{Uzan} indicate that our
universe has positive curvature.

All the above reasons indicate that although people believe in a
flat geometry for the universe, astronomical observations leave
enough room for considering a nonflat geometry. Also about the
interaction between DM and DE there are several signals from
nature which guides us to let the models explain such behaviour.
Based on these motivations we would like here to extend the
studies on GGDE, to a non-flat FRW spacetime in the presence of an
interaction term. Our work differs from \cite{shemov,ebrbd} in
that we consider the GGDE model while in \cite{shemov} and
\cite{ebrbd}, the original GDE model in Einstein and Brans-Dicke
theory were studied, respectively. To check the viability of our
model, we also perform the cosmological constraints on the
interacting GGDE in a non-flat universe by using the Marko Chain
Monte Carlo (MCMC) method. We use the following observational
datasets: Cosmic Microwave Background Radiation (CMB) data from
WMAP7 \cite{wmap7}, $557$ Union2 dataset of type Ia supernova
\cite{Union2}, baryon acoustic oscillation (BAO) data from SDSS
DR7 \cite{sdssbao}, and the cluster X-ray gas mass fraction data
from the Chandra X-ray observations \cite{chandra}. To put the
constraints, we modify the public available CosmoMC
\cite{cosmomc}.

The outline of this paper is as follows. In section III, we study
the cosmological implications of the GGDE scenario in the absence
of interaction between DE and DM. In section \ref{intflat}, we
consider \textit{interacting} GGDE in a flat geometry. In section
\ref{nonflatint}, we generalize the study to the universe with
spacial curvature in the presence of interaction between DM and
DE. In section \ref{constraints}, cosmological constraints on the
parameters of the model are performed by using the Marko Chain
Monte Carlo (MCMC) method. We summarize our results in section
\ref{sum}.
\section{GGDE model in a flat universe}\label{nonintflat}
Consider a flat homogeneous and isotropic FRW universe, the
corresponding Friedmann equation is
\begin{eqnarray}\label{Fried}
H^2=\frac{8\pi G}{3} \left( \rho_m+\rho_D \right),
\end{eqnarray}
where $\rho_m$ and $\rho_D$ are, the energy densities of
pressureless DM and DE, respectively. The generalized ghost energy
density may be written as \cite{caighost2}
\begin{equation}\label{GGDE}
\rho_D=\alpha H+\beta H^2,
\end{equation}
where $\alpha$ is a constant of order $\Lambda_{\rm QCD}^3$ and
$\Lambda_{\rm QCD}$ is QCD mass scale, and $\beta$ is also a
constant. In the original GDE ($\beta=0$) with $\Lambda_{\rm
QCD}\sim 100MeV$ and $H\sim 10^{-33}eV$ , $\Lambda_{\rm QCD}^3 H$
gives the right order of magnitude $\sim (3\times 10^{-3}\rm
{eV})^4$ for the observed DE density \cite{ohta}. In the GGDE,
$\beta$ is a free parameter and can be adjusted for better
agreement with observations.

As usual we introduce the fractional energy density parameters as
\begin{equation}\label{Omega}
\Omega_m=\frac{\rho_m}{\rho_{cr}}= \frac{8\pi G \rho_m}{3 H^2},\ \
\ \Omega_D=\frac{\rho_D}{\rho_{cr}}=\frac{8\pi G(\alpha+\beta
H)}{3H},
\end{equation}
where $\rho_{cr}=3H^2/(8\pi G)$. Thus, we can rewrite the first
Friedmann equation as
\begin{equation}\label{fridomega}
\Omega_m+\Omega_D=1.
\end{equation}
Through this section we consider GGDE in the absence of the
interaction term, thus DE and DM evolves independent of each other
and hence they satisfy the following conservation equations
\begin{eqnarray}
\dot\rho_m+3H\rho_m&=&0,\label{consm}\\
\dot\rho_D+3H\rho_D(1+w_D)&=&0\label{consd}.
\end{eqnarray}
If we take the derivative of relations (\ref{Fried}) and
(\ref{GGDE}) with respect to the cosmic time, we arrive at
\begin{equation}\label{hdot}
\dot{H}=-4\pi G\rho_D(1+u+w_D),
\end{equation}
\begin{equation}\label{dotrho}
\dot{\rho}_D=\dot{H}(\alpha+2\beta H).
\end{equation}
where $u=\rho_m/\rho_D$. Combining relations (\ref{hdot}) and
(\ref{dotrho}) with continuity equation (\ref{consd}), we get
\begin{equation}\label{prewD}
(1+w_D)[3H-4\pi G(\alpha+2\beta H)]=4\pi G (\alpha+2\beta H).
\end{equation}
Solving the above equation for $w_D$  and noticing that
$u={\Omega_m}/{\Omega_D}$, and
\begin{equation}\label{rhomaking}
    \frac{4\pi G}{3 H}(\alpha+2\beta
    H)=\frac{\Omega_D}{2}+\frac{4\pi G \beta}{3},
\end{equation}
we obtain
\begin{equation}\label{wD}
w_D=\frac{\xi-\Omega_D}{\Omega_D(2-\Omega_D-\xi)},
\end{equation}
where $\xi=\frac{8\pi G \beta}{3}$. It is clear that this relation
reduces to its respective one in the GDE when $\xi=0$
\cite{shemov}. In Fig. 1a we have plotted the evolution of $w_D$
versus $\Omega_D$. It is easy to see that at the late time where
$\Omega_D\rightarrow 1$, we have $w_D\rightarrow -1$, which
implies that the GGDE model mimics a cosmological constant
behaviour. One should notice that this behaviour is the same as
for the original GDE model. This is expected since the subleading
term $H^2$ in the late time can be ignored due to the smallness of
$H$ and the difference between these two models appears only at
the early epoches of the universe. From figure (1a) we see that
$w_D$ of the GGDE model cannot cross the phantom divide and the
universe has a de Sitter phase at the late time.
\begin{figure}\epsfysize=5cm
{ \epsfbox{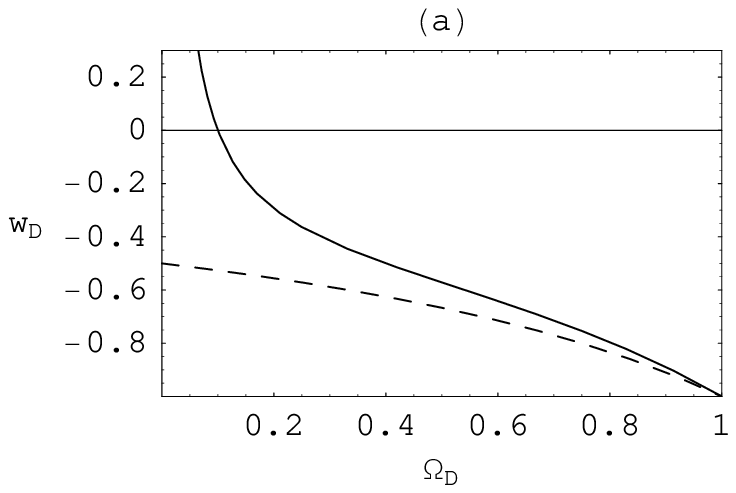}}\epsfysize=5cm {
\epsfbox{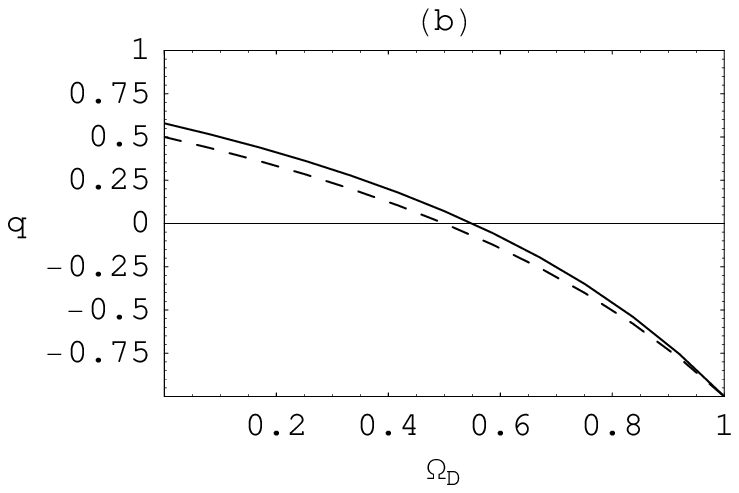}}\caption{These figures show the evolutions
of $w_D$ and $q$ against $\Omega_D$ in a flat GGDE and GDE models.
Solid lines correspond to GGDE when $\xi=0.1$ and the dashed lines
belong to GDE model.} \label{i1}
\end{figure}
It is important to note that the universe is filled with two dark
components namely, DM and GGDE. Thus to discuss the acceleration
of the universe we should define the effective EoS parameter,
$w_{\rm eff}$, as
\begin{equation}\label{wef}
    w_{\rm eff}=\frac{p_t}{\rho_t}=\frac{p_D}{\rho_D+\rho_m},
\end{equation}
where $\rho_t$ and $p_t$ are, respectively, the total energy
density and the total pressure of the universe. As usual, we have
assumed the DM is in the form of pressureless fluid ($p_m$=0).
Using relation (\ref{fridomega}) for the spatially flat universe,
one can find
\begin{equation}\label{wef2}
w_{\rm eff}=\Omega_Dw_D= \frac{\xi-\Omega_D}{2-\Omega_D-\xi}.
\end{equation}
Let us now turn to the deceleration parameter which is defined as
\begin{equation}\label{q1}
q=-\frac{a\ddot{a}}{\dot{a}^2}=-1-\frac{\dot{H}}{H^2},
\end{equation}
where $a$ is the scale factor. Using Eq. (\ref{hdot}) and
definition $\Omega_D$ in (\ref{Omega}) we obtain
\begin{equation}\label{dotH}
\frac{\dot{H}}{H^2}=-\frac{3}{2}\Omega_D \left(1+u+w_D\right).
\end{equation}
Replacing this relation into  (\ref{q1}), and using (\ref{wD}) we
find
\begin{equation}\label{qfnonint}
q=\frac{1}{2}-\frac{3}{2}\frac{\xi-\Omega_D}{(\xi+\Omega_D-2)}.
\end{equation}
One can easily check that the deceleration parameter in GDE is
retrieved for $\xi=0$ \cite{shemov}. We can also take a look at
the early and the late time behaviour of the deceleration
parameter. At the early stage of the universe where
$\Omega_D\rightarrow0$, the deceleration parameter becomes
\begin{equation}\label{earq}
    q=\frac{1}{2}-\frac{3}{2} \frac{\xi}{\xi-2}.
\end{equation}
which indicates that for $\xi<2$ the universe is at the
deceleration phase at early times while for $\xi>2$, the universe
could experience an acceleration phase, the former is consistent
with the definition $\xi=\frac{8\pi G \beta}{3}$. On the other
side, we find that at the late time where the DE dominates
($\Omega_D\rightarrow 1$), independent of the value of the $\xi$,
we have $q=-1$. We have plotted the behaviour of $q$ in Fig.
\ref{i1}b. Besides, taking $\Omega_{D0}=0.72$  and adjusting
$\xi=0.01$ we obtain $q_0 \approx-0.34$, in agreement with
observations \cite{daly}. Choosing the same set of parameters
leads to $w_{D0}\approx -0.78$ and $w_{\rm eff0}\approx -0.56$.
One should note that as we already mentioned about $w_D$, the
squared term in the GGDE density has a negative contribution in
the role of the DE in the universe. We mean by negative
contribution that arises by taking the squared term into account,
the evolution of the universe will be slowed. For example, the
universe will enter the acceleration phase later than the original
GDE. This behaviour is clearly seen in both parts of Fig.
\ref{i1}.

At the end of this section we present the evolution equation of
the DE density parameter $\Omega_D$. To this goal we take the time
derivative of Eq. (\ref{Omega}), after using relation
${\dot{\Omega}_D}= H \frac{d\Omega_D}{d\ln a}$ as well as Eq.
(\ref{q1}) we reach
\begin{equation}\label{Omegaprime1}
\frac{d\Omega_D}{d\ln a}=-3\Omega_D\left(1-\Omega_D\right)w_D.
\end{equation}
Using Eq. (\ref{wD}) we get
\begin{equation}\label{Omegaprime2}
\frac{d\Omega_D}{d\ln a}=-3
\frac{(1-\Omega_D)(\xi-\Omega_D)}{2-\Omega_D-\xi}.
\end{equation}
Once again for the limiting case $\xi=0$, the above relation
reduces to its respective evolution equation for the original GDE
presented in \cite{shemov}.
\section{Interacting GGDE in a flat universe}\label{intflat}
\begin{figure}\epsfysize=5cm
{ \epsfbox{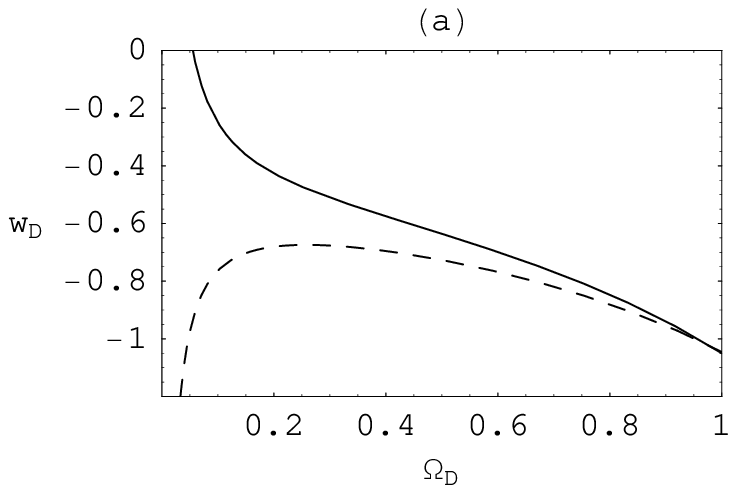}} \epsfysize=5cm {
\epsfbox{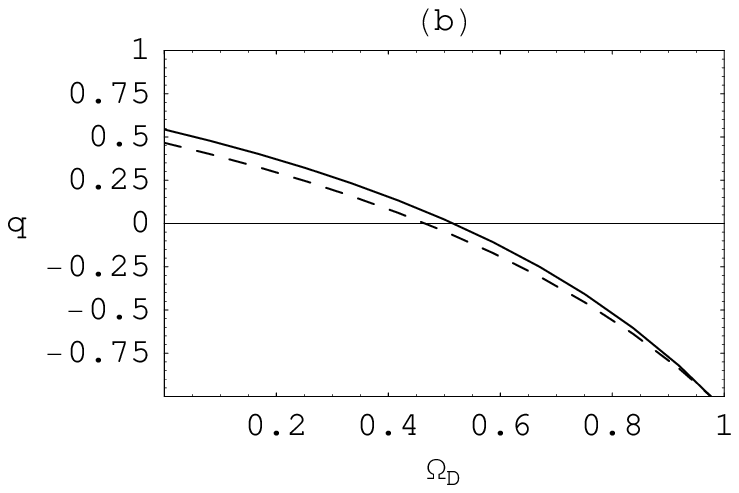}}\caption{These figures show the
evolutions of $w_D$ and $q$ against $\Omega_D$ for a flat
interacting GGDE and GDE model. Solid lines correspond to the GGDE
when $\xi=0.1$ and the dashed lines belong to the GDE model. For
both cases $b=0.15$.} \label{i2}
\end{figure}

In the previous section, the evolution of the DE and DM components
were discussed separately. Here we would like to extend the study
to the interacting case, seeking new features of GGDE. In the
first look investigating interacting models of DE are valuable
from two perspective. The first is the theoretical one, which
states that we have no reason against interaction between DE and
DM components. For example, in the unified models of field theory
DM and DE can be explained by a single scalar field, thus they
will be allowed to interact minimally. Besides, one can get rid of
the coincidence problem by taking into account the interaction
term between DM and DE. One can refer
to\cite{coin1,coin2,coin3,coin4,coin5} for detailed discussion.
The other feature which motivates us to consider interacting
models of DE and DM comes from observations which indicate the
interaction between two dark components of our universe
\cite{interact1}. Thus, there exist enough motivations to consider
the GGDE in the presence of an interaction term. To this end, we
start with the energy balance equations for DE and DM, namely
\begin{eqnarray}
\dot\rho_m+3H\rho_m&=&Q,\label{consmi}\\
\dot\rho_D+3H\rho_D(1+w_D)&=&-Q\label{consdi},
\end{eqnarray}
where $Q>0$ represents the interaction term which allows the
transition of energy from DE to DM. The form of $Q$ is a matter of
choice and can be taken as \cite{shemov}
\begin{equation}\label{Q}
Q =3b^2 H(\rho_m+\rho_D)=3b^2 H\rho_D(1+u),
\end{equation}
with $b^2$  being a coupling constant. Inserting Eqs.
(\ref{dotrho}) and (\ref{Q}) in Eq. (\ref{consdi})  and taking
into account $u=\frac{\Omega_m}{\Omega_D}$, we find
\begin{equation}\label{wD4}
w_D=-\frac{1}{2-\Omega_D-\xi}\left(1+\frac{2b^2}{\Omega_D}-\frac{\xi}{\Omega_D}\right).
\end{equation}
At first look one can find that setting $b=0$, $w_D$ reduces to
the respective relation in the absence of interaction obtained in
Eq. (\ref{wD}). When $\xi=0$ the result recovers those in
\cite{shemov} for original GDE. The first interesting point about
the EoS parameter of the GGDE is that in the interacting case
independent of the interaction parameter, $b$, for $0<\xi<1$,
$w_D$ can cross the phantom line in the future where
$\Omega_D\rightarrow1$. At the present time, by choosing
$\xi=0.03$, $b=0.15$ and $\Omega_{D0}=0.72$, we find that
$w_{D0}=-0.82$ and $w_{\rm eff0}=-0.59$ which the latter favored
by observations. One can easily check that for a same coupling
constant these values for the original GDE are $w_{D0}=-0.83$ and
$w_{\rm eff0}=-0.60$ which clearly show that the square term in
the energy density of the GGDE slow down the evolution of the
universe compared to the original GDE model. For a better insight
we have plotted $w_D$ against $\Omega_D$ in Fig. \ref{i2}a. This
value for coupling constant, $b$, in the figure is consistent with
recent observations \cite{wang1}. It is worth mentioning that at
the late time where $\Omega_D\rightarrow1$ the effective EoS
parameter approaches less than $-1$, i.e. $w_{\rm eff}<-1$, which
reminds a super acceleration for the universe in the future.
Next we take a look at the deceleration parameter in the presence
of an interaction term. Substituting (\ref{dotH}) in (\ref{q1})
and using (\ref{wD4}) yields
\begin{equation}\label{q3}
q=\frac{1}{2}-\frac{3}{2}\frac{\Omega_D}{(2-\Omega_D-\xi)}\left(1+\frac{2b^2}{\Omega_D}-\frac{\xi}{\Omega_D}\right).
\end{equation}
Once again it is clear that setting $b=0$, the respective relation
in the previous section is retrieved. When $\xi=0$ the result of
\cite{shemov} is recovered. For the set of parameters
($\xi=0.03,b=0.15,\Omega_{D0}=0.72$), we find that according to
the GGDE the universe enters the acceleration phase at
$\Omega=0.48$ while this transition happens earlier for the GDE
model. This point is clear from Fig.\ref{i2}b. The present value
of the deceleration parameter for the interacting GGDE model is
$q_0=-0.38$ which is consistent with observations \cite{daly}.

Finally, we would like to obtain the evolution equation of DE in
the presence of interaction. First we take the time derivative of
(\ref{Omega}) and obtain
\begin{equation}\label{omegadot}
\dot{\Omega}=\Omega\left[\frac{\dot{\rho}}{\rho}-2\frac{\dot{H}}{H}\right].
\end{equation}
Using relation (\ref{consdi}) as well as (\ref{dotH}), it is a
matter of calculation to show
\begin{equation}\label{Omegaprime3}
\frac{d\Omega_D}{d\ln a}=3 \Omega_D\left[\frac{1-\Omega_D}{2-
\Omega_D-\xi}\left(1+\frac{2b^2}{\Omega_D}-\frac{\xi}{\Omega_D}\right)-\frac{b^2}{\Omega_D}\right].
\end{equation}
In the limiting case $\xi=0$ the equation of motion of interacting
GDE is recovered \cite{shemov}.
\section{Interacting GGDE in a non-flat universe}\label{nonflatint}
The flatness problem in standard cosmology was resolved by
considering an inflation phase in the evolution history of the
universe. Following this theory it became a general belief that
our universe is spatially flat. However, later it was shown that
exact flatness is not a necessary consequence of inflation if the
number of e-foldings is not very large \cite{huang}. So it is
still possible that there exists a contribution to the Friedmann
equation from the spatial curvature, though much smaller than
other energy components according to observations. Thus,
theoretically the possibility of a curved FRW background is not
rejected. In addition, recent observations support the possibility
of a non-flat universe and detect a small deviation from $k=0$
\cite{nonflat1,nonflat2,nonflat3,nonflat4}. Furthermore, the
parameter $\Omega_k$ represents the contribution to the total
energy density from the spatial curvature and it is constrained as
$-0.0175 <\Omega_k< 0.0085$ with $95\%$ confidence level by
current observations \cite{water}. Our aim in this section is to
study the dynamic evolution of the GGDE in a universe with spatial
curvature. The first Friedmann equation in a non-flat universe is
written as
\begin{eqnarray}\label{Friedm}
H^2+\frac{k}{a^2}=\frac{1}{3M_p^2} \left( \rho_m+\rho_D \right),
\end{eqnarray}
where $k$ is the curvature parameter with $k = -1, 0, 1$
corresponding to open, flat, and closed universes, respectively.
Taking the energy density parameters (\ref{Omega}) into account
and defining the energy density parameter for the curvature term
as $\Omega_k=k/(a^2H^2)$, the Friedmann equation can be rewritten
in the following form
\begin{equation}\label{fridomega2}
1+\Omega_k=\Omega_m+\Omega_D.
\end{equation}
Using the above equation the energy density ratio becomes
\begin{equation}\label{u2}
u=\frac{\rho_m}{\rho_D}=\frac{\Omega_m}{\Omega_D}=\frac{1+\Omega_k-\Omega_D}{\Omega_D}.
\end{equation}
The second Friedmann equation reads
\begin{equation}\label{doth2}
\dot{H}=-4\pi G(p+\rho)+\frac{k}{a^2},
\end{equation}
while the time derivative of GGDE density is
\begin{equation}\label{dotrho3}
\dot{\rho}_D=\dot{H}(\alpha+2\beta H).
\end{equation}
Inserting Eq. (\ref{doth2}) into (\ref{dotrho3}) and combining the
resulting relation with the conservation equation for DE component
(\ref{consdi}), after using (\ref{Q}) and (\ref{u2}), we find the
EoS parameter of interacting GGDE in non-flat universe
\begin{equation}\label{wDn2}
w_D=-\frac{1}{2-\Omega_D-\xi}\left(2-\left(1+\frac{\xi}{\Omega_D}\right)\left(1+\frac{\Omega_k}{3}\right)+\frac{2b^2}{
\Omega_D} (1+\Omega_k)\right).
\end{equation}
\begin{figure}\epsfysize=5cm
{ \epsfbox{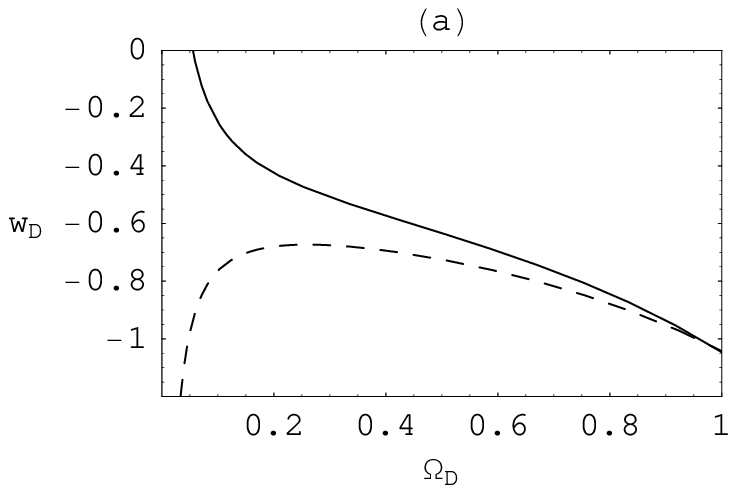}} \epsfysize=5cm {
\epsfbox{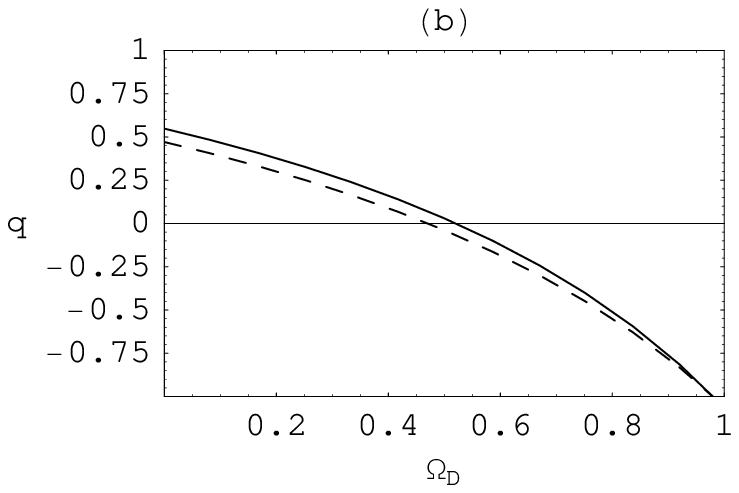}}\caption{These figures show
 the evolutions of $w_D$ and $q$ against $\Omega_D$ for a
interacting GGDE and GDE models in a non-flat universe. Solid
lines correspond to the GGDE when $\xi=0.1$ and the dashed lines
belong to the GDE model. For both cases $b=0.15$.} \label{i3}
\end{figure}
From the second Friedmann equation, (\ref{doth2}), one can easily
obtain
\begin{equation}\label{q1n}
\frac{\dot{H}}{H^2}=-\Omega_k+\frac{3 }{2} \Omega_D[1+u+w_D],
\end{equation}
and therefore the deceleration parameter in a non-flat background
is obtained as
\begin{equation}\label{q1n}
q=-1-\frac{\dot{H}}{H^2}=-1-\Omega_k+\frac{3 }{2}
\Omega_D[1+u+w_D].
\end{equation}
Substituting Eqs. (\ref{u2}) and (\ref{wDn2}) in (\ref{q1n}) we
obtain
\begin{equation}\label{q2n}
q=\frac{1}{2}\left(1+\Omega_k\right)-\frac{3\Omega_D}{2(2-\Omega_D-\xi)}
\left[2-\left(1+\frac{\xi}{\Omega_D}\right)\left(1+\frac{\Omega_k}{3}\right)+\frac{2b^2}{
\Omega_D} (1+\Omega_k)\right].
\end{equation}
In a non-flat FRW universe, the equation of motion of interacting
GGDE is obtained following the method of the previous section. The
result is
\begin{equation}\label{Omegaprime2n}
\frac{d\Omega_D}{d\ln
a}=3\Omega_D\left[\frac{\Omega_k}{3}+\frac{1-\Omega_D}{2-\Omega_D-\xi}
\left(2-\left(1+\frac{\xi}{\Omega_D}\right)\left(1+\frac{\Omega_k}{3}\right)+\frac{2b^2}{
\Omega_D}
(1+\Omega_k)\right)-\frac{b^2}{\Omega_D}(1+\Omega_k)\right].
\end{equation}
In the limiting case $\Omega_k=0$, the results of this section
restore their respective equations in a flat FRW universe derived
in the previous sections, while for $\xi=0$ the respective
relations in \cite{shemov} are retrieved. The evolutions of $w_D$
and $q$ against $\Omega_D$ for a non-flat interacting GGDE and GDE
models are plotted in Fig.\ref{i3}. Let us explore different
features of GGDE in non-flat universe by a numerical study. First
of all we study the EoS parameter of the GGDE in the future where
$\Omega_D\rightarrow1$. In this case, taking $\xi=0.1$, $b=0.15$
and $\Omega_k=0.01$ leads to $w_{D}=-1.05$ which indicates that
the GGDE is capable to cross the phantom line in the future. The
present stage of the universe can be achieved by the same set of
parameters but $\Omega_D=0.72$. In such a case we see that
$w_{D0}=-0.78$ while the effective EoS parameter becomes $w_{\rm
eff0}=-0.6$ which is consistent with observations. The
deceleration parameter of the model can also be obtained which is
in agreement with observational evidences. For example, for the
above choice of parameters one finds $q_0=-0.34$ \cite{daly}.
Transition from deceleration to the acceleration phase, in the
interacting non-flat case, take place at $\Omega_D=0.52$.

\section{Cosmological Constraints}\label{constraints}
In order to constrain our model parameters space and check its
viability, we apply the Marcov Chain Monte Carlo (MCMC) method.
Observational constraints on the original GDE with and without
bulk viscosity, was already performed \cite{Obs}. Our work differs
from \cite{Obs} in that we consider the GGDE with energy density
$\rho_D= \alpha H+\beta H^2$, while the authors of \cite{Obs}
studied the original GDE with energy density $\rho_D= \alpha H$.
Besides, we have extended here the study to the universe with any
spacial curvature. To make a fitting on the cosmological
parameters the public available CosmoMC package \cite{cosmomc} has
been modified.

\subsection{Method}\label{method}
We want to get the best value of the parameters with $1\sigma$
error at least. Thus, following \cite{Obs}, we employ the maximum
likelihood method where the total likelihood function
$\mathcal{L}=e^{-\chi^2/2}$ is the product of the separate
likelihood functions
\begin{equation}
 \chi^2_{tot}=\chi^2_{SNIa}+\chi^2_{CMB}+\chi^{2}_{BAO}+\chi^2_{gas}.\label{totchi1}
\end{equation}
Here SNIa stands for type Ia supernova, BAO for baryon acoustic
oscillation and gas stands for $X$-ray gas mass fraction data. The
best fitting values of parameters are obtained by minimizing
$\chi_{tot}^2$. In the next subsection, every dataset will be
discussed separately.

We employ the following datasets. CMB data from WMAP7
\cite{wmap7}, 557 Union2 dataset of type Ia supernova
\cite{Union2}, baryon acoustic oscillation (BAO) data from SDSS
DR7 \cite{sdssbao}, and the cluster X-ray gas mass fraction data
from the Chandra X-ray observations datasets \cite{chandra}.

\subsubsection{Cosmic Microwave Background}
For the CMB data, we use the WMAP7 dataset \cite{wmap7}. The shift
parameter R, which parametrize the changes in the amplitude of the
acoustic peaks is given by \cite{bond}
\begin{equation}
 R=\sqrt{\frac{\Omega_{m0}}{c}}\int_{0}^{z_{\ast}}\frac{dz^{\prime}}{E(z^{\prime})},\label{RCMB}
\end{equation}
where $z_{\ast}$ is the redshift of decoupling. In addition, the acoustic scale $l_{A}$,
which characterizes the changes of the peaks of CMB via the angular diameter distance out to the decoupling is
defined as well in \cite{bond} by
\begin{equation}
 l_{A}=\frac{\pi r(z_{\ast})}{r_{s}(z_{\ast})}\label{lCMB}.
\end{equation}
The comoving distance $r(z)$ is defined
\begin{equation}
 r(z)=\frac{c}{H_{0}}\int_{0}^{z}\frac{dz^{\prime}}{E(z^{\prime})},\label{comdis}
\end{equation}
and  the comoving sound horizon  at the recombination $r_{s}(z_{\ast})$ is written
\begin{equation}
 r_{s}(z_{\ast})=\int_{0}^{a(z_{\ast})}\frac{c_{s}(a)}{a^2H(a)}da,\label{shdis}
\end{equation}
and the sound speed $c_{s}(a)$ is defined by
\begin{equation}
 c_{s}(a)=\left[3(1+\frac{3\Omega_{b0}}{4\Omega{\gamma0}}a)\right]^{-1/2},\label{soundspeed}
\end{equation}
where the seven-year WMAP observations gives $\Omega_{\gamma 0}=2.469\times10^{-5}h^{-2}$ \cite{wmap7}.
\par The redshift $z_{\ast}$ is obtained by using the fitting function proposed by Hu and Sugiyama \cite{Hu}
\begin{equation}
 z_{\ast}=1048[1+0.00124(\Omega_{b0}h^2)^{-0.738}][1+g_{1}(\Omega_{m0}h^2)^{g_{2}}],\label{redshift}
\end{equation}
where
\begin{equation}
 g_1=\frac{0.0783(\Omega_{b0}h^2)^{-0.238}}{1+39.5(\Omega_{b0}h^2)^{0.763}}, \hspace{1cm} g_2=\frac{0.560}{1+21.1(\Omega_{b0}h^2)^{1.81}},\label{g1g2}
\end{equation}

\par Then one can define  $\chi^2_{CMB}$  as $\chi^2_{CMB}=X^TC^{-1}_{CMB}X$, with \cite{Obs,wmap7}
\begin{subequations}
\begin{align}
 X&=\begin{pmatrix} l_{A}-302.09 \\R-1.725\\z_{\ast}-1091.3 \end{pmatrix},\label{CCMB},\\
 C^{-1}_{CMB}&=\begin{pmatrix} 2.305 & 29.698 &-1.333 \\293689 & 6825.270 & -113.180\\-1.333 & -113.180 & 3.414 \end{pmatrix},\label{invcovCMB}
 \end{align}
\end{subequations}
where $C^{-1}_{CMB}$ is the inverse covariant matrix.
\subsubsection{Type Ia Supernovae Data}
We shall use the SNIa Union2 dataset \cite{Union2} which includes
$577$ SNIa. The Hubble parameter $H(z)$ determines the history of
the universe. However, $H(z)$ is specified by the underlying
theory of gravity. To test this model, we can use the
observational data for some predictable cosmological parameter
such as luminosity distance $d_{L}$. One may note that the Hubble
parameter $H(z;\alpha_{1},...,\alpha_{n})$ can describe the
universe, where  parameters $(\alpha_1,...\alpha_n)$ are predicted
by the cosmological model. For such a cosmological model we can
define the theoretical 'Hubble-constant free' luminosity distance
as
\begin{equation}
 D^{th}_{L}=H_{0}\frac{d_{L}}{c}=(1+z)\int_{0}^{z}\frac{dz^{\prime}}{E(z^{\prime};\alpha_z,...,\alpha_n)}=H_{0}\frac{1+z}{\sqrt{|\Omega_{k}|}}\mathrm{sinn}\left[\sqrt{|\Omega_{k}|}
\int_{0}^{z}\frac{dz^{\prime}}{H(z^{\prime};\alpha_z,...,\alpha_n)}\right],\label{LD}
\end{equation}
where $E\equiv\frac{H}{H_{0}}$, $z$ is the redshift parameter, and

\[ \mathrm{sinn}(\sqrt{|\Omega_{k}|}x) = \left\{ \begin{array}{ll}
         \sin(\sqrt{|\Omega_{k}|}x) & \mbox{for $\Omega_{k} < 0$}\\
    \sqrt{|\Omega_{k}|}x & \mbox{for $\Omega_{k} = 0$}\\
        \sinh(\sqrt{|\Omega_{k}|}x) & \mbox{for $\Omega_k > 0$}.\end{array} \right. \] \label{OmegaK}

Then one can write the theoretical modulus distance
\begin{equation}
 \mu_{th}(z)=5\log_{10}[D_{L}(z)]+\mu_{0}\; ,\label{muth}
\end{equation}
where $\mu_{0}=5\log_{10}(cH_{0}^{-1}/Mpc)+25$. On the other hand,
the observational modulus distance of the SNIa, $\mu_{obs}(z_i)$,
at redshift $z_i$ is given by
\begin{equation}
 \mu_{obs}(z_i)=m_{obs}(z_i)-M,\label{muobs}
\end{equation}
where m and M are apparent and absolute magnitudes of SNIa respectively.
Then the parameters of the theoretical model, $\alpha_{i}$s, can be determined
by a likelihood analysis by defining $\chi^2_{SNIa}(\alpha_i, M^{\prime})$ in Eq. (\ref{totchi1}) as
\begin{eqnarray}
 \chi^2_{SNIa}(\alpha_i, M^{\prime})&\equiv&\sum_{j}\frac{(\mu_{obs}(z_j)-\mu_{th}(\alpha_{i},z_j))^2}{\sigma_{j}^2}\\ \nonumber
&=&\sum_{j}\frac{(5\log_{10}[D_{L}(\alpha_{i},z_j)]-m_{obs}(z_j)+M^{\prime})^2}{\sigma_{j}^2}\;
,
\end{eqnarray}
where the nuisance parameter, $M^{\prime}=\mu_{0}+M$, can be marginalized over as
\begin{equation}
 \bar{\chi}^{2}_{SNIa}(\alpha_{i})=-2\ln\int_{-\infty}^{+\infty}\exp[-\frac{1}{2}\chi^2_SN(\alpha_i, M^{\prime})]dM^{\prime}\; .
\end{equation}

\subsubsection{Baryon Acoustic Oscillation}
The baryon acoustic oscillations data from the Sloan Digital Sky Survey
(SDSS) Data Release 7 (DR7) \cite{sdssbao} is used here for
constraining the model parameters. The data constrains
$d_{z}\equiv r_{s}(z_{d})/D_{V}(z)$, where $r_{s}(z_{d})$ is the comoving sound
 horizon at the drag epoch (where baryons were released from photons) and $D_{V}$ is given by \cite{eisen1}
\begin{equation}
D_{V}(z)\equiv\left[ \left(\int_{0}^{z}\frac{dz^{\prime}}{H(z^{\prime})}\right)^2\frac{cz}{H(z)}\right]^{1/3}\; ,\label{DBAO}
\end{equation}
The drag redshift is given by the fitting formula \cite{eisen2}
\begin{equation}
 z_{d}=\frac{1291(\Omega_{m0}h^2)^{0.251}}{1+0.659(\Omega_{m0}h^2)^{0.828}}\left[1+b_{1}(\Omega_{b0}h^2)^{b_{2}}\right]\; ,\label{dragredshift}
\end{equation}
where
\begin{equation}
 b_{1}=0.313(\Omega_{m0}h^2)^{-0.419}[1+0.607(\Omega_{m0}h^2)^{0.607}], \hspace{1cm} b_{2}=0.238(\Omega_{m0}h^2)^{0.223}\; . \label{b1b2bao}
\end{equation}
Then we can obtain $\chi^2_{BAO}$ by $\chi^2_{BAO}=Y^TC^{-1}_{BAO}Y$, where
\begin{equation}
 Y=
\begin{pmatrix}
 d_{0.2}-0.1905\\d_{0.35}-0.1097\end{pmatrix}\; ,\label{YBAO}
\end{equation}
 and its covariance matrix is given by \cite{sdssbao}
\begin{equation}
 C^{-1}_{BAO}=\begin{pmatrix}30124 & -17227\\-17227 & 86977\end{pmatrix}\; .\label{CBAO}
\end{equation}
These results are similar to those obtained in \cite{Obs} for
original GDE in flat universe.
\subsubsection{X-Ray Gas Mass Fraction}

The ratio of the X-ray gas mass to the total mass of a cluster is defined
as  X-ray gas mass fraction  \cite{chandra}.
The $\Lambda$CDM model proposed \cite{chandra}
\begin{eqnarray}
f_{gas}^{\Lambda CDM}(z)=\frac{K A \gamma b(z)}{1+s(z)}\left(\frac{\Omega_b}{\Omega_{0m}}\right)
\left(\frac{D_A^{\Lambda CDM}(z)}{D_A(z)}\right)^{1.5}\; .
\label{fgasLCDM}
\end{eqnarray}
The elements in Eq. (\ref{fgasLCDM}) are defined as follows:
 $D_{A}^{\Lambda CDM} (z)$ and $D_{A}(z)$ are
the proper angular diameter distance in the $\Lambda$CDM
and the interested model respectively. Angular correction
factor $A$
\begin{eqnarray}
A=\left(\frac{\theta_{2500}^{\Lambda CDM}}{\theta_{2500}}\right)^\eta \approx
\left(\frac{H(z)D_A(z)}{[H(z)D_A(z)]^{\Lambda CDM}}\right)^\eta\; ,
\end{eqnarray}
is caused by the change in angle for the our interested
model $\theta_{2500}$ in comparison with $\theta_{2500}^{\Lambda CDM}$,
where $\eta=0.214\pm0.022$ \cite{chandra} is the slope of the $f_{gas}(r/r_{2500})$
data within the radius $r_{2500}$.
 The proper angular diameter distance is given by
\begin{eqnarray}
D_A(z)=\frac{c}{(1+z)\sqrt{|\Omega_k|}}\mathrm{sinn}\left[\sqrt{|\Omega_k|}\int_0^z\frac{dz'}{H(z')}\right]\;
.
\end{eqnarray}
The bias factor $b(z)$ in Eq.  (\ref{fgasLCDM}) contains
information about the uncertainties in the cluster depletion
factor $b(z)= b_0(1+\alpha_b z)$, the parameter $\gamma$ accounts
for departures from the hydrostatic equilibrium. The function
$s(z)=s_0(1 +\alpha_s z)$ denotes the uncertainties of the
baryonic mass fraction in stars with a Gaussian prior for $s_0$,
with $s_0=(0.16\pm0.05)h_{70}^{0.5}$ \cite{chandra}. The factor
$K$ describes the combined effects of the residual uncertainties,
such as the instrumental calibration, and a Gaussian prior for the
'calibration' factor is considered by $K=1.0\pm0.1$
\cite{chandra}.
\par Then, $\chi^2_{gas}$ is defined as \cite{chandra}
\begin{eqnarray}
&&\chi^2_{gas}=\sum_i^N\frac{[f_{gas}^{\Lambda
CDM}(z_i)-f_{gas}(z_i)]^2}{\sigma_{f_{gas}}^2(z_i)}+\frac{(s_{0}-0.16)^{2}}{0.0016^{2}}
+\frac{(K-1.0)^{2}}{0.01^{2}}+\frac{(\eta-0.214)^{2}}{0.022^{2}}\ ;,\label{chifgas}
\end{eqnarray}
with the statistical uncertainties $\sigma_{f_{gas}}(z_{i})$.

\subsection{Results}\label{resultssec}
Finally, the maximum likelihood method is applied for the
interacting GGDE in a non-flat universe by using the CosmoMc  code
\cite{cosmomc}. Figure. \ref{resfig1} shows 2-D contours with
$1\sigma$ and $2\sigma$ confidence levels where 1-D distribution
of the model parameters are shown as well. Best fit parameter
values are shown in Table. \ref{results1} with
 $1\sigma$ and $2\sigma$ confidence levels. From Table \ref{results1}
we can see that the best fit results are given as:
$\Omega_{0DE}=0.7145^{+0.0427+0.0484}_{-0.0264-0.0452} $,
$\Omega_{0m}= 0.2854^{+0.0264+0.0452}_{-0.0427-0.0467}$,
$\Omega_{0k}=0.0285^{+0.0014}_{-0.0274}$. In addition for the
model parameters the best fit values are obtained as:
$\xi=0.2300^{+0.4769}_{-0.0129}$,  $b=0.0592^{+0.1407}_{-0.0492}$.
The age of the universe in this model is given by
$13.7385^{+0.3302+0.3796}_{-0.2907-0.3313}$ Gyr. We have also
plotted the  evolution of $\omega_D$, $\Omega_D$ and $q$ against
the scale factor $a$ for the interacting GGDE in a nonflat
universe by using the best fit values of the model parameters.
\begin{table}
    \begin{tabular}{|c||c|c|}
        \hline
        Parameter     & Best Fit value & Mean Value\\ \hline\hline
        $\Omega_{b}h^2$ & ~   $0.0226^{+0.0016+0.0015}_{-0.0016-0.0021}$           & ~  $0.02257^{+0.0005+0.0009}_{-0.0004-0.0010}$   \\ \hline
        $\Omega_{DM}h^2 $& ~    $0.1153^{+0.0061+0.0099}_{-0.0106-0.0119}$          & ~   $0.1132^{+0.0030+0.0063}_{-0.0030-0.0058}$  \\ \hline
        $\Omega_{0m}$   & ~   $0.2854^{+0.0264+0.0452}_{-0.0427-0.0467}$           & ~  $0.2769^{+0.0129+0.0284}_{-0.0131-0.0242}$   \\ \hline
        $\Omega_{0k}$   & ~   $0.0285^{+0.0014}_{-0.0274}$           & ~  $0.0187^{+0.0112}_{-0.0117}$   \\ \hline
        $\Omega_{0DE}$  & ~      $0.7145^{+0.0427+0.0484}_{-0.0264-0.0452}$        & ~  $0.7230^{+0.0131+0.0242}_{-0.0129-0.0284}$   \\ \hline
    $b$         & ~       $0.0592^{+0.1407}_{-0.0492}$       & ~   $0.1082^{+0.0917}_{-0.0982}$  \\ \hline
    $\xi$   & ~     $0.2300^{+0.4769}_{-0.0129}$        & ~ $0.2228_{+0.2771}^{-0.2128}$ \\ \hline
        $H_{0}$         & ~      $69.5401^{+3.5998+4.2626}_{-2.6037-3.5376}$        & ~  $70.0610_{-1.1138-2.3955}^{+1.1566+2.2773}$   \\ \hline
        Age (Gyr)       & ~      $13.7385^{+0.3302+0.3796}_{-0.2907-0.3313}$        & ~ $13.7596_{-0.1065-0.2102}^{+0.1072+0.2173}$    \\ \hline
    \end{tabular}
\caption{The best fit and mean values of the  model parameter with $1\sigma$ and $2\sigma$ regions  from MCMC calculation by using CMB, SNIa
 Union2, X-gas and BAO datasets.
The Hubble parameter is in the unit of $km s^{-1}Mpc^{-1}$.} \label{results1}
\end{table}
\begin{figure}\epsfysize=5cm
{\epsfbox{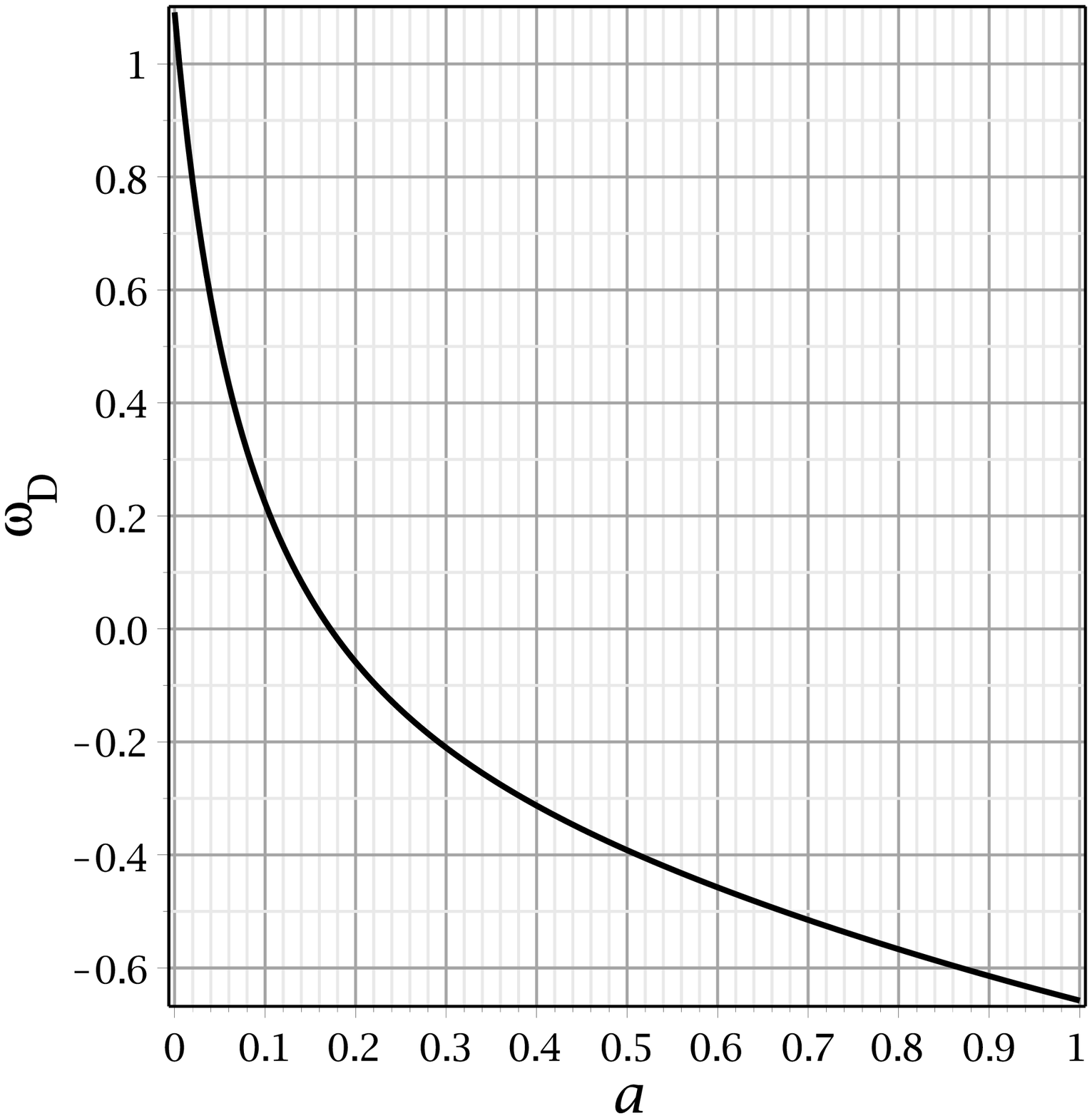}} \epsfysize=5cm {
\epsfbox{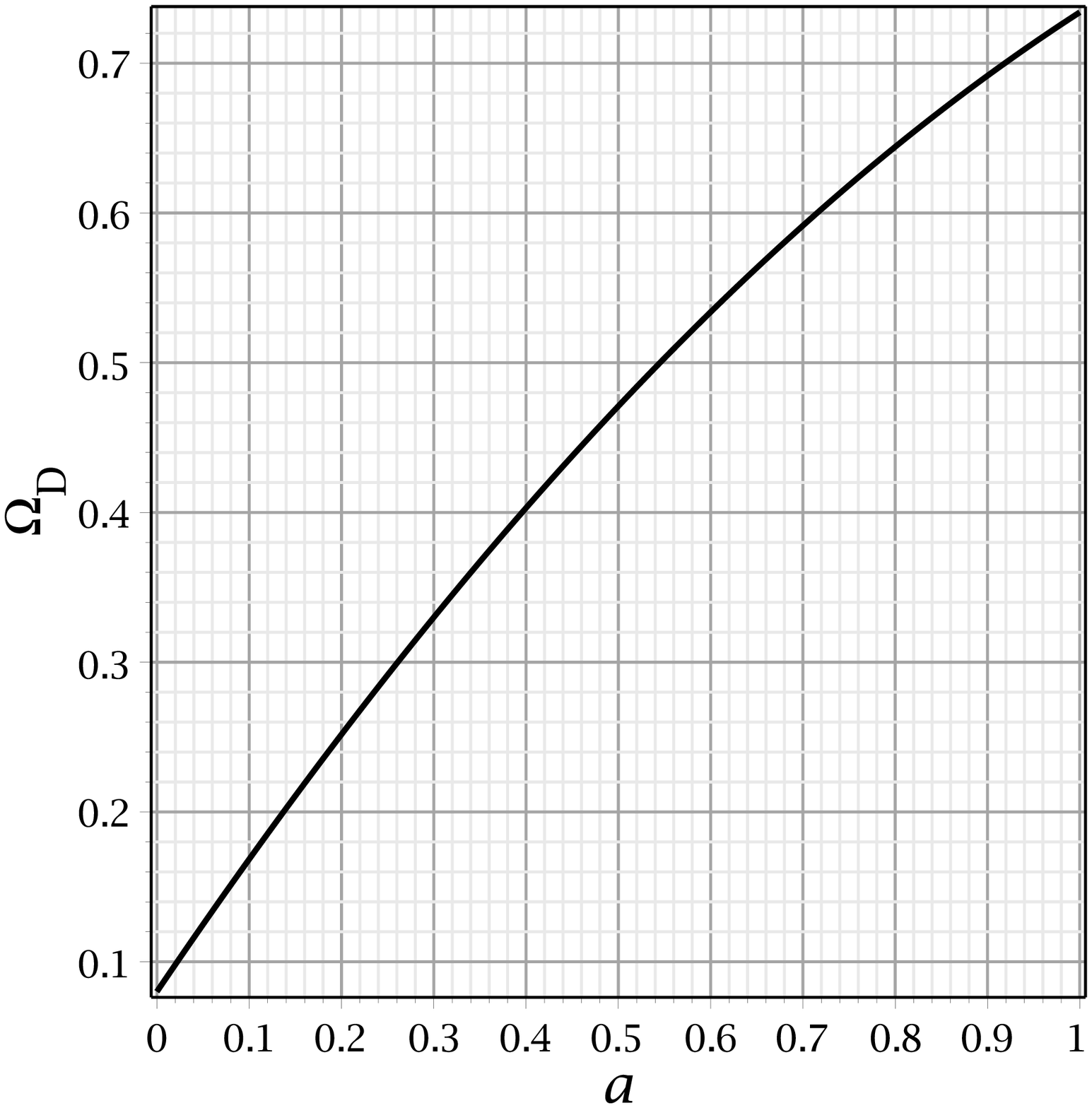}}\epsfysize=5cm {
\epsfbox{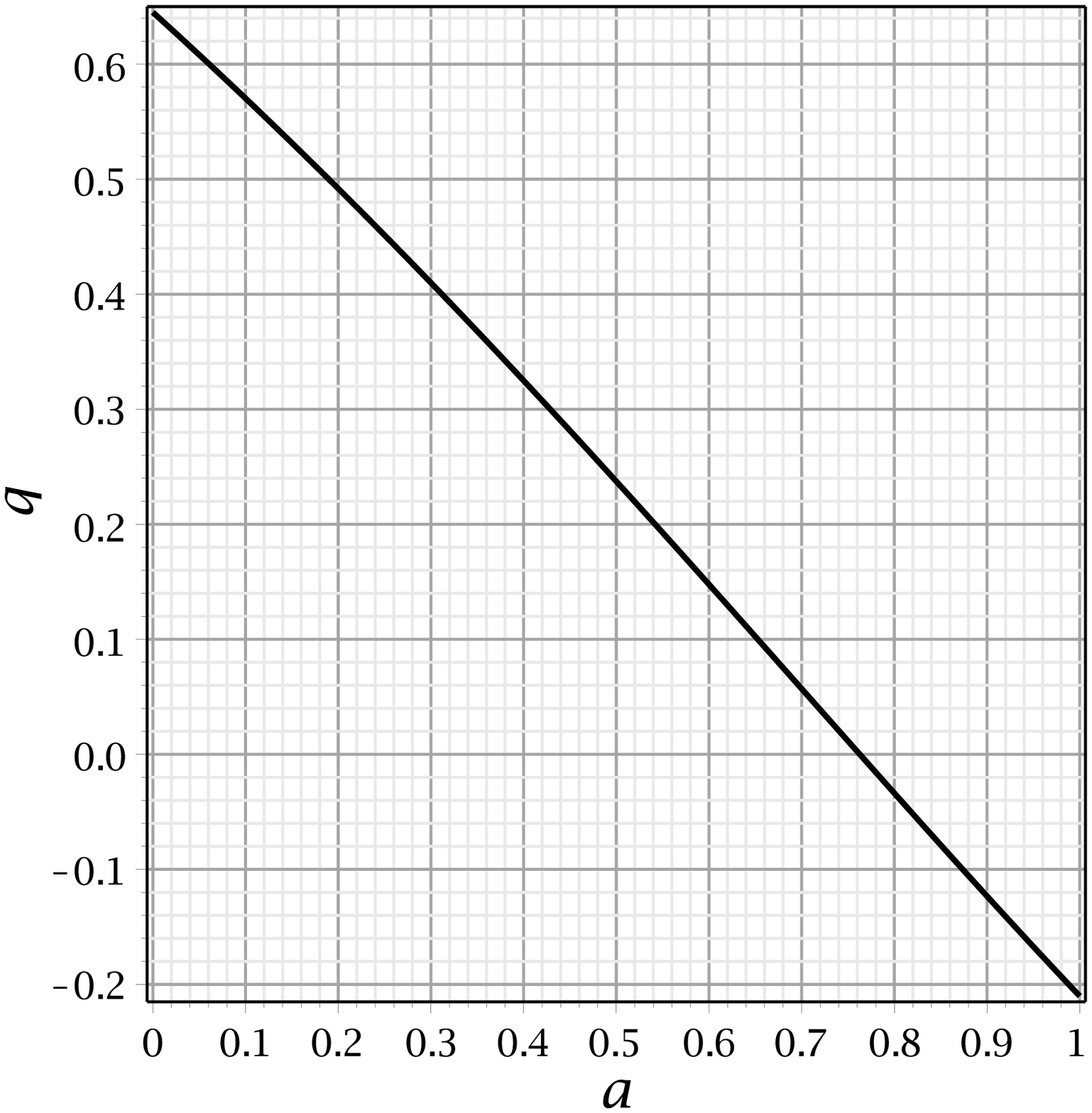}}\caption{These figures show
 the evolutions of $w_D$, $\Omega_D$ and $q$  against  the scale factor $a$ for the
interacting GGDE models in a nonflat universe, where $\xi=0.23$,
$b=0.05$ and $\Omega_k=0.028$ which are chosen from the best fit
values of Table \ref{results1}.} \label{bestfigplot}
\end{figure}

\begin{figure}[htb]
        \center{\includegraphics[width=\textwidth]
        {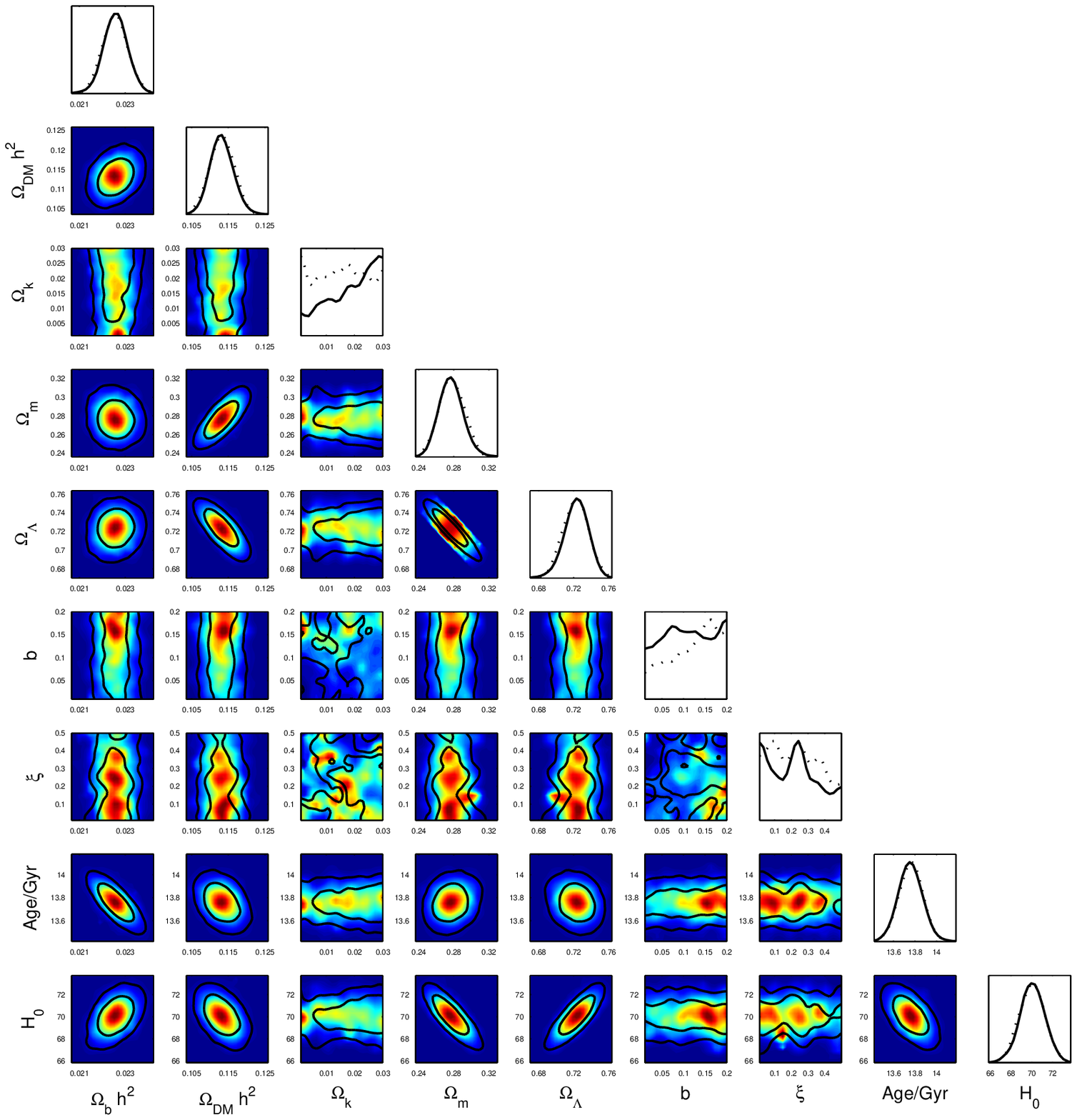}}
        \caption{ 1-D constraints on parameters and their 2-D contours
with $1\sigma$ and $2\sigma$ regions. To obtain these plots,
Union2+CMB+BAO+X-gas with BBN constraints are used. In 1-D plots, the solid lines are mean likelihoods of samples and dotted
lines are marginalized probabilities for each parameter.}\label{resfig1}
      \end{figure}

\section{Summary and discussion}\label{sum}
In order to resolve the DE puzzle, people usually prefer to handle
the problem by using existing degree's of freedom. GDE is a
prototype of these models which discusses the acceleration of the
universe and originates from vacuum energy of the Veneziano ghost
field in QCD. This model can address the fine tuning problem
\cite{shemov}. An extended version of this model called GGDE was
recently proposed by Cai et. al., \cite{caighost2}, seeking a
better agreement with observations.

In this paper we explored some features of GGDE in both flat and
non-flat FRW universe in the presence of an interaction term
between the two dark components of the universe. In section II, we
discussed the GGDE in a flat FRW background. We found that the EoS
parameter approaches $-1$ which is the same as the cosmological
constant. The next section was devoted to the interacting GGDE in
a flat geometry. An interesting feature which we found was the
capability of crossing the phantom line in this case. This
behaviour is also seen in the last section for interacting GGDE in
a universe with spatial curvature.

Then, we applied the Markov Chain Monte Carlo method together with
the latest observational data to constrain the model parameters.
The results are presented in Table \ref{results1} and Fig.
\ref{resfig1}. The main result found through this paper is that in
the GGDE model, there is a delay in different epoches of the
cosmic evolution in comparison with original GDE model. This
result was also pointed out in \cite{caighost2} due to the
negative contribution of the square term in the energy density of
GGDE.
\acknowledgments{We are grateful to the referees for constructive
comments which helped us to improve the paper significantly. A.
Sheykhi thanks from the Research Council of Shiraz University.
This work has been supported financially by Research Institute for
Astronomy \& Astrophysics of Maragha (RIAAM) Iran.}


\begin{thebibliography}{99}
\bibitem{Rie} A.G. Riess, et al., Astron. J.  116 (1998)
1009;\\
  S. Perlmutter, et al.,  Astrophys. J.  517 (1999) 565;\\
  S. Perlmutter, et al.,  Astrophys. J.  598 (2003) 102;\\
  P. de Bernardis, et al.,  Nature  404 (2000) 955.

\bibitem{sami} E. J. Copeland, M. Sami and S. Tsujikawa, Int. J. Mod. Phys. D
15, 1753 (2006).

\bibitem{li} M. Li, X. -D. Li, S. Wang, Y. Wang, Commun. Theor. Phys. 56,
525-604 (2011).

\bibitem{dgp} G. R. Dvali, G. Gabadadze and M. Porrati, Phys. Lett. B 485 (2000) 208.

\bibitem{deffayet} C. Deffayet,  Phys. Lett. B 502 (2001) 199.

\bibitem{arkani1} N. Arkani Hamed, H. C. Cheng, M. A. Luty and S. Mukohyama, JHEP 05 (2004) 074.

\bibitem{arkani2} N. Arkani Hamed, H. C. Cheng, M.A. Luty, S. Mukohyama and T. Wiseman, JHEP 01 (2007) 036.

\bibitem{Nozari} K. Nozari, S. D. Sadatian, Eur. Phys. J. C {58}, 499
(2008);\\ K. Nozari, B. Fazlpour, JCAP {0806}, 032 (2008).

\bibitem{Cor} C. Corda, Int. J. Mod. Phys. D {18}, 2275 (2009).
\bibitem{sheykhi} A. Sheykhi, Phys. Lett. B 680 (2009) 113;\\  
A. Sheykhi, Class. Quantum Gravit. 27 (2010) 025007;\\ 
A. Sheykhi, Phys. Lett. B681 (2009) 205;\\  
K. Karami, et. al., Gen. Relativ. Gravit. 43 (2011) 27;\\ 
M. Jamil, A. Sheykhi, Int. J. Theor. Phys. 50 (2011) 625;\\ 
A. Sheykhi, M. Jamil, Phys. Lett. B 694 (2011) 284. 

\bibitem{witten} E. Witten, Nucl. Phys. B 156 (1979) 269;\\  G. Veneziano,
Nucl. Phys. B 159 (1979) 213;\\ C. Rosenzweig, J. Schechter and C.
G. Trahern, Phys. Rev. D 21 (1980) 3388.


\bibitem{urban1} F. R. Urban and A. R. Zhitnitsky, Phys. Lett. B 688 (2010)
9;\\ K.~Kawarabayashi and N.~Ohta, Nucl.\ Phys.\ B {175} (1980)
477.

\bibitem{ohta} N. Ohta, Phys. Lett. B 695 (2011) 41,
arXiv:1010.1339.

\bibitem{caighost} R.G. Cai, Z.L. Tuo, H.B. Zhang, arXiv:1011.3212.

\bibitem{shemov} A. Sheykhi, M.Sadegh Movahed, Gen Relativ Gravit 44 (2012)
449. 

\bibitem{ebrins} E. Ebrahimi and A. Sheykhi, Int. J. Mod. Phys. D 20 (2011) 2369.

\bibitem{shemoveb} A. Sheykhi, M. Sadegh Movahed, E. Ebrahimi, Astrophys Space
Sci 339 (2012) 93.

\bibitem{shebagh} A. Sheykhi, A. Bagheri, Euro. Phys. Lett., 95 (2011)
39001.

\bibitem{ebrbd} E. Ebrahimi and A. Sheykhi, Phys. Lett. B 706 (2011)
19.

\bibitem{rozas} A. Rozas-Fernandez, Phys. Lett. B 709 (2012)
313.

\bibitem{karami} K. Karami, M. Mousivand, arXiv:1209.2044.

\bibitem{khodam1} A. Khodam-Mohammadi, M. Malekjani, M.
Monshizadeh, Mod. Phys. Lett. A 27, 18 (2012) 1250100.

\bibitem{khodam2} M. Malekjani, A. Khodam-Mohammadi,
arXiv:1202.4154.


\bibitem{Obs} Chao-Jun Feng, Xin-Zhou Li, Xian-Yong Shen, Phys. Rev. D 87 (2013)
023006.


\bibitem{Zhi} A. R. Zhitnitsky, arXiv:1112.3365.

\bibitem{mag} M. Maggiore, L. Hollenstein, M. Jaccard and E. Mitsou, Phys. Lett.
B 704, 102 (2011).

\bibitem{caighost2} R. G. Cai, Z. L. Tuo, Y. B. Wu, Y. Y. Zhao,     Phys.Rev. D86 (2012) 023511. 

\bibitem{interact1} O. Bertolami , F. Gil Pedro and M. Le Delliou, Phys. Lett. B
654 (2007) 165.

\bibitem{oli} G. Olivares, F. Atrio, D. Pavon, Phys. Rev. D 71 (2005) 063523.

\bibitem{guth}  A. H. Guth, Phys. Rev. {D 23},347 (1981).

\bibitem{spe} C. L. Bennett, et al.,  Astrophys. J. Suppl.
148 (2003) 1;\\ D. N. Spergel, Astrophys. J. Suppl. 148 (2003) 175;\\
M. Tegmark, et al., Phys. Rev. D 69 (2004) 103501;\\ U. Seljak, A.
Slosar, P. McDonald, JCAP 0610 (2006) 014;\\ D. N. Spergel, et
al., Astrophys. J. Suppl. 170 (2007) 377.


\bibitem{Sie} J. L. Sievers, et al., Astrophys. J. 591 (2003) 599;\\ C.B.
Netterfield, et al., Astrophys. J. 571 (2002) 604;\\ A. Benoit, et
al., Astron. Astrophys. 399 (2003) L25;\\ A. Benoit, et al.,
Astron. Astrophys. 399 (2003) L19.

\bibitem{Caldwell} R. R. Caldwell, M. Kamionkowski, astro-ph/0403003;\\ B. Wang,
Y. G. Gong, R. K. Su, Phys. Lett. B 605 (2005) 9.


\bibitem{Uzan} J. P. Uzan, U. Kirchner, G.F.R. Ellis, Mon. Not. R. Astron.
Soc. 344 (2003) L65;\\ A. Linde, JCAP 0305 (2003) 002;\\ M.
Tegmark, A. de Oliveira-Costa, A. Hamilton, Phys. Rev. D 68 (2003)
123523;\\ G. Efstathiou, Mon. Not. R. Astron. Soc. 343 (2003)
L95;\\ J. P. Luminet, J. Weeks, A. Riazuelo, R. Lehou, J. Uzan,
Nature 425 (2003) 593; \\ G. F. R. Ellis, R. Maartens, Class.
Quantum Grav. 21 (2004) 223.



\bibitem{wmap7} E. Komatsu et al., Astrophys. J. Suppl. 192, 18 (2011).
\bibitem{Union2} R. Amanullah, et al., Astrophys. J. 716, 712 (2010).
\bibitem{sdssbao} B. A. Reid et al., Mon. Not. Roy. Astron. Soc. 401, 2148 (2010).

\bibitem{chandra} S. W. Allen, et al., Mon. Not. Roy. Atsron. Soc. 383 879 (2008).
\bibitem{cosmomc} A. Lewis and S. Bridle, Phys. Rev. D 66, 103511 (2002).

\bibitem{daly} R.A. Daly, et al., J. Astrophys. 677 (2008) 1.


\bibitem{coin1} L. Amendola, Phys. Rev. D 62, 043511 (2000);\\ L. Amen- dola and
C. Quercellini, Phys. Rev. D 68 (2003) 023514 ;\\ L. Amendola, S.
Tsujikawa and M. Sami, Phys. Lett. B 632 (2006) 155.

\bibitem{coin2} D. Pavon,W. Zimdahl, Phys. Lett. B 628 (2005) 206;\\ S. Campo,
R. Herrera, D. Pavon, Phys. Rev. D 78 (2008) 021302(R).

\bibitem{coin3} C. G. Boehmer, G. Caldera-Cabral, R. Lazkoz, R. Maartens,
Phys. Rev. D 78 (2008) 023505.

\bibitem{coin4} G. Olivares, F. Atrio-Barandela and D. Pavon, Phys.
Rev. D 74 (2006) 043521.

\bibitem{coin5} S. B. Chen, B. Wang, J. L. Jing, Phys.Rev. D 78 (2008) 123503.

\bibitem{wang1} B. Wang, Y. Gong and E. Abdalla, Phys. Lett. B 624
(2005) 141;\\ B. Wang, C. Y. Lin and E. Abdalla, Phys. Lett. B 637
(2005) 357.

\bibitem{huang} Q. G. Huang and M. Li, J. Cosmol. Astropart. Phys.
JCAP08 (2004)013.

\bibitem{nonflat1} C. L. Bennett, et al., Astrophys. J. Suppl. 148 (2003) 1;\\ D.
N. Spergel, Astrophys. J. Suppl. 148 (2003) 175;\\ M. Tegmark, et
al., Phys. Rev. D 69 (2004) 103501;\\ U. Seljak, A. Slosar, P.
McDonald, JCAP 0610 (2006) 014;\\ D. N. Spergel, et al.,
Astrophys. J. Suppl. 170 (2007) 377.

\bibitem{nonflat2} J. L. Sievers, et al., Astrophys. J. 591 (2003) 599;\\ C.B.
Netterfield, et al., Astrophys. J. 571 (2002) 604;\\ A. Benoit, et
al., Astron. Astrophys. 399 (2003) L25;\\ A. Benoit, et al.,
Astron. Astrophys. 399 (2003) L19.

\bibitem{nonflat3} R. R. Caldwell, M. Kamionkowski, JCAP 0409 (2004) 009;\\
B. Wang, Y. G. Gong, R. K. Su, Phys. Lett. B 605 (2005) 9.

\bibitem{nonflat4} J. P. Uzan, U. Kirchner, G.F.R. Ellis, Mon. Not. R.
Astron. Soc. 344 (2003) L65;\\ A. Linde, JCAP 0305 (2003) 002;\\
M. Tegmark, A. de Oliveira-Costa, A. Hamilton, Phys. Rev. D 68
(2003) 123523;\\ G. Efstathiou, Mon. Not. R. Astron. Soc. 343
(2003) L95;\\ J. P. Luminet, J. Weeks, A. Riazuelo, R. Lehou, J.
Uzan, Nature 425 (2003) 593;\\ G. F. R. Ellis, R. Maartens, Class.
Quantum Grav. 21 (2004) 223.

\bibitem{water} T. P. Waterhouse and J. P. Zipin, arXiv:0804.1771.


\bibitem{bond} J. R. Bond, et al, Mon. Not. Roy. Astron. Soc. 291, L33 (1997).
\bibitem{Hu} W. Hu and N. Sugiyama, Astrophys. J. 471, 452 (1996).
\bibitem{eisen1}  D. J. Eisenstein et al., Astrophys. J. 633, 560 (2005).
\bibitem{eisen2} D. J. Eisenstein and W. Hu, Astrophys. J. 496 605 (1998).

\end{thebibliography}
\end{document}